\documentclass[a4paper,12pt]{spieman}  
\usepackage{amsmath,amsfonts,amssymb}
\usepackage{graphicx}
\usepackage{setspace}
\usepackage{tocloft}
\usepackage{gensymb}
\usepackage{xcolor}
\usepackage{soul}
\usepackage[left]{lineno}

\title{Rapid characterization of exoplanet atmospheres with the Exoplanet Transmission Spectroscopy Imager (ETSI)}

\author[abcd]{Luke M. Schmidt}
\author[abce]{Ryan J. Oelkers}
\author[abc]{Erika Cook}
\author[f]{Mary Anne Limbach}
\author[abc]{D. L. DePoy}
\author[abc]{J. L. Marshall}
\author[ac]{Landon Holcomb}
\author[ac]{Willians Pena}
\author[gac]{Jacob Purcell}
\author[ac]{Enrique Gonzalez Vega}

\affil[a]{Department of Physics and Astronomy Texas A\&M University, 4242 TAMU, College Station, TX 77843-4242 USA}
\affil[b]{George P. and Cynthia Woods Mitchell Institute for Fundamental Physics and Astronomy, Texas A\&M University, College Station, TX, 77843-4242 USA}
\affil[c]{Charles R. \& Judith G. Munnerlyn Astronomical Laboratory, Department of Physics \& Astronomy, Texas A\&M University, College Station, TX 77843, USA}
\affil[d]{Yerkes Observatory, 373 W. Geneva Street, Williams Bay, WI 53191 USA}
\affil[e]{Department of Physics and Astronomy, The University of Texas, Rio Grande Valley, Brownsville, TX 78520 USA}
\affil[f]{Department of Astronomy,University of Michigan, Ann Arbor, MI 48109 USA}
\affil[g]{Department of Physics, Indiana University - Purdue University at Indianapolis (IUPUI), 402 North Blackford Street, Indianapolis, Indiana 46202-3273}

\cftpagenumbersoff{figure}
\cftpagenumbersoff{table} 
\begin{document} 
\maketitle

\begin{abstract}
The Exoplanet Transmission Spectroscopy Imager (ETSI) amalgamates a low resolution slitless prism spectrometer with custom multi-band filters to simultaneously image 15 spectral bandpasses between 430 nm and 975 nm with an average spectral resolution of $R = \lambda/\delta\lambda \sim 20$. ETSI requires only moderate telescope apertures ($\sim2$ m) and is capable of characterizing an exoplanet atmosphere in as little as a single transit, enabling selection of the most interesting targets for further characterization with other ground and space-based observatories and is also well suited to multi-band observations of other variable and transient objects. This enables a new technique, common-path multi-band imaging (CMI), used to observe transmission spectra of exoplanets transiting bright (V$<$14 magnitude) stars. ETSI is capable of near photon-limited observations, with a systematic noise floor on par with the Hubble Space Telescope and below the Earth's atmospheric amplitude scintillation noise limit. We report the as-built instrument optical and optomechanical design, detectors, control system, telescope hardware and software interfaces, and data reduction pipeline. A summary of ETSI's science capabilities and initial results are also included.
\end{abstract}


{\noindent \footnotesize\textbf{*}Luke Schmidt,  \linkable{lschmidt@yerkesobservatory.org} }

\section{INTRODUCTION}
\label{sec:intro}  
The advantages of simultaneous multi-band photometry were leveraged by astronomers in the mid 20th century. Early experiments with photomultiplier tubes by Walraven resulted in the Walraven five channel spectrophotometer\cite{1960BAN....15...67W} and along with other instruments such as the Danish $uvby-\beta$ five channel spectrophotometer\cite{1987Msngr..50...45F} demonstrated that simultaneous measurement of multiple bands is superior to sequential observations of multiple bands when measuring relative colors of astronomical objects. Examples of simultaneous multi-band photometry observations include stellar pulsations of subluminous B (sdB) stars\cite{Falter:2003}, constraining physical properties of transiting exoplanets\cite{10.1093/mnras/stv2698}, changes in the spectral energy distribution of supernovae \cite{2024arXiv240508327Y} and characterization of spots on M-dwarf stars\cite{10.1093/mnras/stae841}. Changes in seeing and transparency can still affect the recorded brightness of an object, but any variable absorption (e.g. clouds), which is colorless, has the same effect on all wavelengths \cite{EverettHowell:2001, Ivezic:2007}, leaving relative color measurements unaffected if the measurements are made simultaneously. The first multi-band instruments were complex and required very stable environments as the bandpass selection was performed by slicing a dispersed spectrum with a set of prisms. Any flexure or motion of the star in the instrument field of view could change the bandpass seen in a given channel.  Other multi-band instruments involved multiple telescopes observing with different filters\cite{10.1117/12.2055167} or multiple dichroics, filters, and detectors, one for each bandpass \cite{10.1093/mnras/stv197, 10.1117/1.JATIS.5.1.015001, 10.1117/12.927094, 10.1117/12.2233643, 2007MNRAS.378..825D, 10.1117/12.2054933, 10.1117/12.2504817}, which become unwieldy beyond only a few bandpasses due to mechanical space constraints as well as transmission losses for channels at the end of the chain of stacked dichroics. Fortunately, recent advances in thin film coating technology \cite{johansen2016} have enabled the creation of filters with many bandpasses selected by a single optic and when used in both transmission and reflection enable simultaneous measurement of many more bandpasses with only two detectors.

In this paper we describe the instrument design, construction, and commissioning activities as well as a brief description of the wide variety of objects and types of observations that are well matched to ETSI's capabilities.  We also highlight the benefits of using ETSI as a reconnaissance instrument to optimize observations with the James Webb Space Telescope\cite{2006SSRv..123..485G} (JWST), or as a rapid follow-up instrument for transient alerts from the Vera C. Rubin Observatory Legacy Survey of Space and Time (LSST)\cite{2019ApJ...873..111I}.

\section{THE ETSI INSTRUMENT}
\begin{figure}
    \centering
    \includegraphics[width=\textwidth]{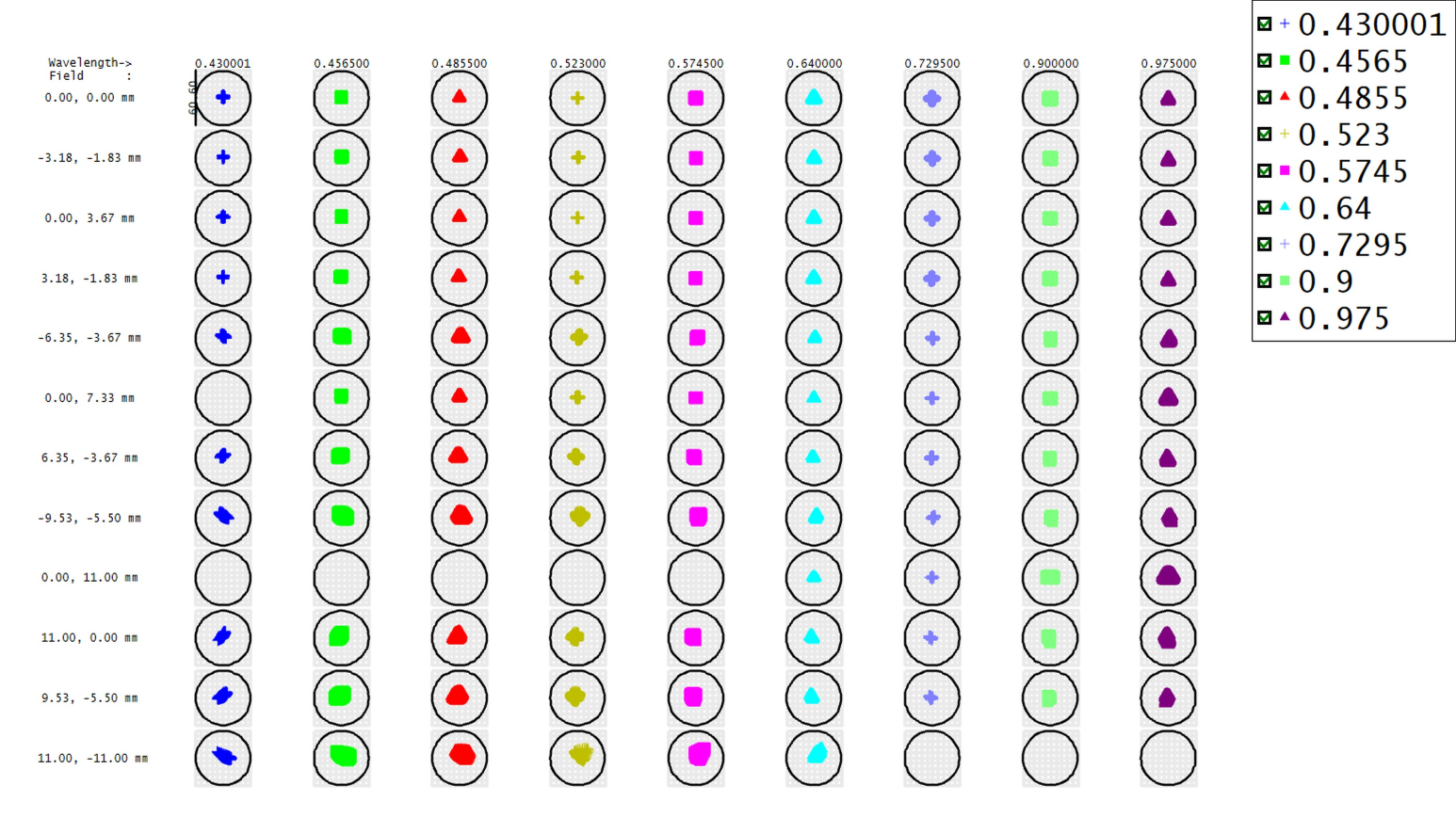}
    \caption{\small{Matrix spot diagram of ETSI and a paraxial McDonald Observatory 2.1 m telescope. The circles are 1 arcsecond in diameter.  The field positions with missing spots are at the edge of the field and those wavelengths fall off the edge of the detector (2048 x 2048, 11$\mu m$ pixels).}}\label{fig:etsi-spot-diagram} 
\end{figure}

The Exoplanet Transmission Spectroscopy Imager (ETSI)\cite{Limbach:2020, Schmidt2022} is based on traditional slitless spectroscopy, with a novel multi-band filter used to split a continuous spectrum into multiple distinct spectral bands, enabling traditional point spread function (PSF) fitting or aperture photometry of each band. This is more efficient and uncomplicated compared to other exoplanet transmission spectroscopy data reduction pipelines (examples in \cite{2014Natur.505...69K, 2024arXiv240520361A}). The complexities of these pipelines, which can include binning higher resolution spectra, wavelength calibration, and spectral fitting techniques, may introduce biases to the retrieved information\cite{10.1093/mnras/stae1073}. The common optical path and simultaneous nature of ETSI observations mean that any instrumental or environmental common path errors, including atmospheric scintillation, are greatly reduced. By referencing one science target spectral band to another science target band, color changes on the order of 0.006\% over the course of a transit are detectable, (see section \ref{sec:data-reduction}). The ETSI optical design was optimized to ensure the point source $80\%$ encircled energy at all wavelengths is well below typical seeing (1-2 arcseconds) and that the PSF's would not appreciably change shape for small changes in focus (see Fig. \ref{fig:etsi-spot-diagram} for wavelength dependent spot diagram and Table \ref{tab:broadband-compliance} for as built performance). This minimizes residual color errors due to wavelength-dependent PSF changes during an observation.

\begin{table}[ht]
\caption{Collimator and cameras as-built optical performance. Requirement was on-axis 80\% encircled energy within a $25.4\mu m$ spot size for the collimator and on-axis 80\% encircled energy within a $12.7\mu m$ spot size for the cameras, tested at three wavelengths.  The cameras and collimator were tested separately, rather than in combination, as that did not require the vendor to develop a custom test bench.}
\label{tab:broadband-compliance}
\begin{center}       
\begin{tabular}{|l|c|c|c|c|}
\hline
\rule[-1ex]{0pt}{3.5ex}  Filter & Field Position & Collimator EE80 ($\mu m$) & Camera 1 EE80 ($\mu m$) & Camera 2 EE80 ($\mu m$)  \\
\hline
\rule[-1ex]{0pt}{3.5ex}  440 nm & on axis & 5.2 & 7.6 & 4.2  \\
\hline
\rule[-1ex]{0pt}{3.5ex}  632 nm & on axis & 11.5 & 5.1 & 9.4 \\
\hline
\rule[-1ex]{0pt}{3.5ex}  900 nm & on axis & 7.6 & 6.0 & 7.9 \\
\hline 
\end{tabular}
\end{center}
\end{table}

ETSI does not measure absolute transit depth in each spectral band, instead it measures relative transit depth between each band. Once an image sequence is obtained on-sky using this instrument, we use a self-referencing differential photometric technique to eliminate sources of uncertainty from non-common path sources such as atmospheric scintillation, instrumental effects, and telescopic effects. To do this we ratio each spectral band of the science star to another spectral band of the same science star (see Equation~\ref{eqn:etsi_comp} below), which differs from traditional photometric measurements made by ratioing the flux of the science target to one or more reference stars (see Equation~\ref{eqn:trad_comp} below). 

\begin{equation}
    f(target)_{i} = f(target)'_{i} / f(target)'_{j}
    \label{eqn:etsi_comp}
\end{equation}

where $f(target)_{i}$ is the detrended flux for the $i^{th}$ bandpass, $f(target)'_{i}$ is the raw flux from the $i^{th}$ bandpass and $f(target)'_{j}$ is the raw flux from the reference bandpass, $j$.

\begin{equation}
    f(target)_{i} = f(target)'_{i} / f(comp)_{i}
    \label{eqn:trad_comp}
\end{equation}

where $f(target)_{i}$ is the detrended flux for the target star in the $i^{th}$ bandpass, $f(target)'_{i}$ is the raw flux from the target star in the $i^{th}$ bandpass, and $f(comp)_{i}$ is the raw flux from the comparison star in the $i^{th}$ bandpass.
Our initial on-sky results show that the highest precision measurements are obtained by using a spectral band near the middle of the ETSI spectral range as our reference band. ETSI color precision is typically limited by photon noise from the target or background even for very bright targets, demonstrating that the systematic errors are well removed. This self-referential photometry eliminates nearly all common-path systematics and allows for a theoretical differential photometric precision on the order of $10-25$~ppm \cite{Limbach:2022}. 

\section{OPTICAL DESIGN}
Typically, exoplanet transmission spectroscopy is conducted using higher-resolution spectroscopic instruments ($R=\lambda/\delta \lambda$ of hundreds to thousands) on large ground and space-based observatories. These instruments are not specialized for the specific task of obtaining exoplanet transmission spectra, and are, perhaps unsurprisingly, sub-optimal for the task.  Specifically, there is typically substantial room for improvement in the instrument transmission and systematic errors. Data reduction processes can be greatly eased (thereby decreasing systematic errors) by building an instrument that produces isolated, binned spectral measurements rather than needing to extract, calibrate, and bin high spectral resolution data. 

ETSI is a purpose-built instrument designed specifically to reduce systematic errors and produce high photometric precision exoplanet transmission spectra. The requirements considered throughout the optical design process are listed below.
\begin{enumerate}
    \item Optimization wavelengths 430-975 nm: Wavelength coverage was driven by coverage of modeled exoplanet atmospheric features and constrained by the sensitivity of the ETSI detectors.
    \item Compatible with $f/8$ or slower telescopes: We envisioned ETSI traveling to multiple telescopes (for example, both Northern and Southern hemisphere telescopes) and this ensured compatibility with a variety of potential telescopes. Currently, ETSI has only been paired with the McDonald Observatory 2.1 m telescope as it has sufficient aperture for exoplanet targets, is easy to interface with, and had a sufficient number of available nights for commissioning and science observations.
    \item Moderate focal reduction: The dispersed nature of the ETSI images mean that each band covers a fairly large number of pixels, which would be exacerbated by use at any telescope with a large focal ratio.  Focal reduction also increases the field of view to increase the likelihood of comparison stars which may be useful during data reduction.
    \item Imaging quality sufficient to ensure seeing limited performance: This also ensures that any residual color, or minor focus changes affect the PSF of each filter band $\sim$equally.
    \item Maintain the angle of the multi-chroic at $30\degree$ or less: driven by the difficulty of designing the multiband filters at higher angles of incidence.
    \item Maintain at least 20 mm of distance between the final camera optical element and the focal plane: This is a reasonable distance that was compatible with the majority of currently available detectors that were considered.
\end{enumerate}
Additional considerations for each optical assembly are described in the following sections. 
\begin{figure}
    \centering
    \includegraphics[width=\textwidth]{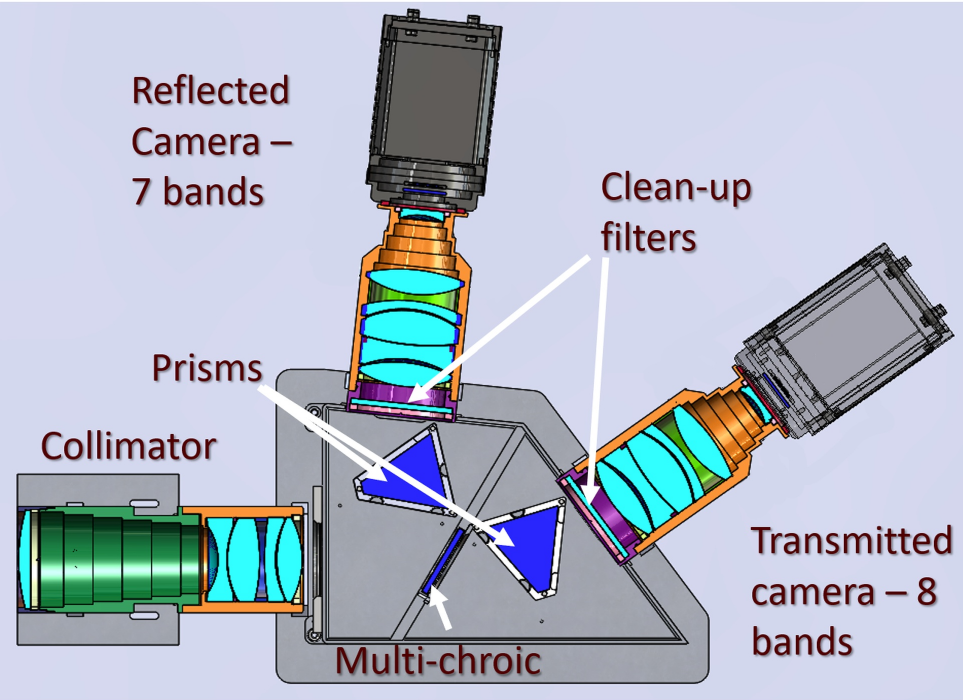}
    \caption{\small{Section view of the ETSI optics and detectors.  The telescope focal plane is to the left of the image.  Light enters the collimator and is split into two channels by the multi-chroic which is located at the pupil.  Identical prisms disperse the light which is then further filtered to sharpen the transmission cut on/off transitions and imaged with identical cameras onto sCMOS detectors.}}\label{fig:etsi-layout} 
\end{figure}

\subsection{Collimator \& Cameras}
ETSI has a five-element collimator and two identical six-element cameras as shown in Figure \ref{fig:etsi-layout}.  The optical prescription is given in Table \ref{tab:optical-prescription} in the Appendix. The collimator is a reverse telephoto with the five lenses separated into three groups.   Much of the design effort was focused on extending the distance from the last element of the collimator to the pupil.  This was required to accommodate the $30\degree$ angle of the multi-chroic to ensure there was adequate mechanical space for the prism in the reflected channel, and that the collimator and camera lens barrels would have adequate clearance as well as space for a shutter to facilitate capturing calibration frames.  

The six-element cameras are split into three groups, with the final element being a field flattener. The camera optimization was constrained by the requirements listed above, as well as the required mechanical spacing between the pupil and front lens to accommodate the prisms and clean-up filters. 

All lens elements for both collimator and cameras are air-spaced with spherical surfaces. The camera and collimator were first developed separately, then the majority of the optimization was as a complete system in order to set the collimator-prism-camera angles as well as tune the prisms to achieve the desired spacing between filter bands (the simulated spacing between bands ranges between 402$\mu m$ and 754$\mu m$ or 36.5-68.5 pixels for 11$\mu m$ pixels). The band spacing was fairly conservative to account for the possibility of bad seeing and/or the potential for filter transmission/reflection transitions that were not as steep as desired. In the end, the filters that were produced were of excellent quality and the prism dispersion could have been reduced.

Final optimization of the optical design involved extensive discussion with the optical manufacturer (JML Optical) about glass blank availability and constraints on lens sizes as well as optimization of glass choices for a minimum of axial color. Of particular use was a technique where all air and glass thicknesses and surface radii were held fixed and only glass substitution was enabled for a limited catalog of glass types.  The glass types were based on vendor availability, with some exclusions due to cost or manufacturing concerns. The merit function weights used in the original optical optimization were almost entirely set to zero for the glass substitution optimization.  Constraints remained on the focal length and axial color, and final optimization minimized the root mean square change in focal length at each of the ETSI filter bands \cite{Siew:16}.

Manufacturing tolerances were determined for each lens as well as positional tolerances for both individual elements and optical sub-assemblies (collimator lens barrel, prism, camera lens barrel, filters). Scratch/dig surface quality tolerance was 20-10 to ensure minimal scattered light.

In combination with the McDonald Observatory 2.1 m telescope ($f/13.7$ at Cassegrain focus), ETSI has an overall focal ratio of $f/6$, which results in a plate scale of $0.18"/pixel$ and a $6.2'$ field of view. ETSI re-images a 72 mm diameter field to a 32 mm diameter image circle.

\subsection{Filters}
The key optical components that make ETSI possible are the multi-chroic and clean-up filters in each channel. These filters are thin film optical interference filters with many alternating layers of high and low index dielectric materials. Filters are generally deposited using Physical Vapor Deposition (PVD) in vacuum chambers and layers
are controlled to within a few nanometers of physical thickness. A complete description of the filters and the filter design process is given in Limbach et al.\cite{10.1117/12.2562371}. The multi-chroic is located at the pupil and is responsible for the inital splitting of the light into eight transmitted bands and seven reflected bands.  The $30\degree$ angle of the multi-chroic and the out of band blocking (Optical Density 3 req. Optical Density 4 goal) and transmission to reflection transition zones made it difficult to meet specifications with only a single optic, so each camera has a clean up filter to improve out of band blocking and cut-on/off sharpness for each of the fifteen filter bands. 

\begin{figure}
    \centering
    \includegraphics[width=\textwidth]{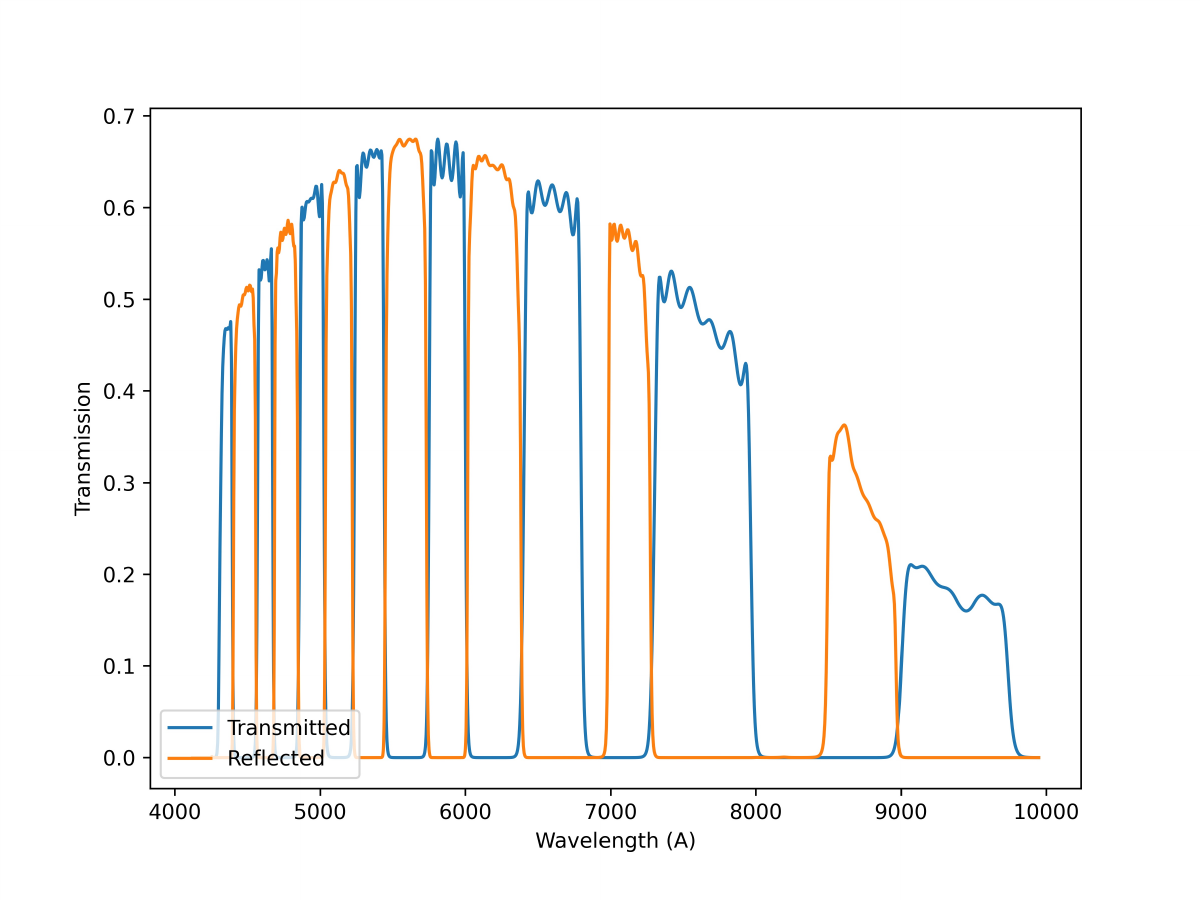}
    \caption{\small{The efficiency of both channels including collimator, multi-chroic, prisms, clean up filters, camera optics, and detector quantum efficiency from vendor measurements. The gap in coverage at $\sim 6900\ \text{\AA}$ and $\sim 8250\ \text{\AA}$} is to ensure sufficient separation between bands on the detector when dispersed.}\label{fig:etsi-transmission} 
\end{figure}

The filter bands were chosen to coincide with exoplanet atmospheric features. We modeled dozens of exoplanet atmospheres using Exo-Transmit\cite{2017PASP..129d4402K} and aligned the spectral bands with detectable molecular and atomic absorption features\cite{10.1117/12.2562371}. Two small portions of the spectrum (680-697 nm and 797-847 nm) are blocked completely in order to maintain adequate spacing between spectral bands of interest on the detectors.  Figure \ref{fig:etsi-transmission} shows the combined transmission of the ETSI optical components, separated into reflected and transmitted bands. The filter bands are listed in Table \ref{tab:filter-bands}

\begin{table}[ht]
\caption{Selected ETSI filter bands.}
\label{tab:filter-bands}
\begin{center}       
\begin{tabular}{|l|c|c|c|c|}
\hline
\rule[-1ex]{0pt}{3.5ex}  Band & [R]eflected or [T]ransmitted & $\lambda$ on - $\lambda$ off [nm] & Target Feature  \\
\hline
\rule[-1ex]{0pt}{3.5ex}  1 & T & 430-440 & Rayleigh  \\
\hline
\rule[-1ex]{0pt}{3.5ex}  2 & R & 440-456.5 & Rayleigh \\
\hline
\rule[-1ex]{0pt}{3.5ex}  3 & T & 456.5-468 & Rayleigh \\
\hline 
\rule[-1ex]{0pt}{3.5ex}  4 & R & 468-485.5 & Rayleigh \\
\hline
\rule[-1ex]{0pt}{3.5ex}  5 & T & 485.5-503 & Rayleigh \\
\hline
\rule[-1ex]{0pt}{3.5ex}  6 & R & 503-523 & Reference \\
\hline
\rule[-1ex]{0pt}{3.5ex}  7 & T & 523-544 & Reference \\
\hline
\rule[-1ex]{0pt}{3.5ex}  8 & R & 544-574.5 & Na \\
\hline
\rule[-1ex]{0pt}{3.5ex}  9 & T & 574.5-600.5 & Na \\
\hline
\rule[-1ex]{0pt}{3.5ex}  10 & R & 600.5-640 & Na \\
\hline
\rule[-1ex]{0pt}{3.5ex}  11 & T & 640-680 & Reference \\
\hline
\rule[-1ex]{0pt}{3.5ex}  12 & R & 697-729.5 & Reference \\
\hline
\rule[-1ex]{0pt}{3.5ex}  13 & T & 729.5-797 & K \\
\hline
\rule[-1ex]{0pt}{3.5ex}  14 & R & 847-900 & CH$_{4}$ \\
\hline
\rule[-1ex]{0pt}{3.5ex}  15 & T & 900-975 & H$_{2}$O \\
\hline
\end{tabular}
\end{center}
\end{table}

ETSI measures fifteen simultaneous bandpasses, but in principle two (one reflected, one transmitted) to $\sim$fifty (25 reflected, 25 transmitted) bands are possible with current technology.  Larger numbers of bands require the bands to be spaced equally in wave number to maximize the filter efficiency and achieve the required cut-on/off transmission slope.

\subsection{Prisms}
\begin{figure}
    \centering
    \includegraphics[width=0.7\textwidth]{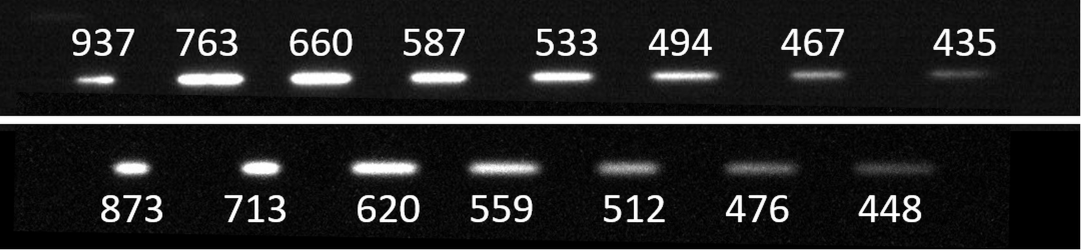}
    \caption{\small{Images of  an F-type star showing the 8 bandpasses from the transmitted channel (top) and 7 bandpasses from the reflected channel (bottom), offset to show that when combined, ETSI captures almost complete wavelength coverage from 430-975 nm.  Bands are red to blue, left to right.  Central wavelength is given in nm above/below each band.}}\label{fig:sample-etsi-images} 
\end{figure}

In order to perform photometry on each individual band, prisms (N-SF5 glass type) disperse the light, separating the filter bands into distinct PSFs which results in the unique dashed-line appearance of ETSI images (see Figure \ref{fig:sample-etsi-images}).

\begin{figure}
    \centering
    \includegraphics[width=12cm]{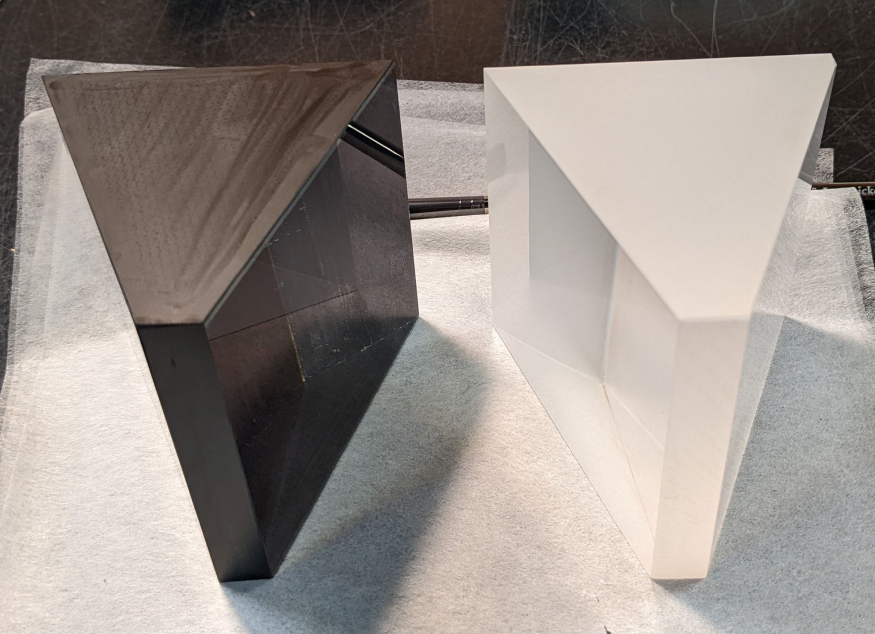}
    \caption{\small{The ETSI prisms, the left prism has had the non-optical surfaces painted with Speedball Super Black India Ink to reduce scattered light.}}\label{fig:prisms} 
\end{figure}

The optical face of each prism is 75 mm on a side with a prism angle of $46.24\degree$.  A complete discussion of the prism design process is described in Limbach et al.\cite{10.1117/12.2562371}.  Attempts were made to develop a zero-deviation prism design, but were ultimately discarded due to the number of prism elements and materials required, and space constraints. The blue bands are slightly over-dispersed and produce a lower signal-to-noise ratio (SNR), especially on cooler stars. However, multiple bands in the blue portion of the spectrum can be combined in post-processing in cases in which the SNR is insufficient to provide useful exoplanet atmospheric constraints. Gratings were not considered due to scattered light concerns, very low required groove densities, and lower transmission when compared to prisms. Note that high instrument transmission is extremely important for exoplanet transmission spectroscopy as the exoplanet atmosphere signal is faint compared to that of the star and sufficient photons must be collected to ensure that the planetary atmosphere signal is above the photon noise of the host star.  The final prism angle and position relative to the collimator and cameras was included in the overall optical design optimization.
To further minimize stray and scattered light, the non-optical surfaces of each prism were blackened with Speedball Super Black India Ink, chosen for its low reflectivity across the full optical spectrum and ease of application\cite{10.1117/12.2562759}. Figure \ref{fig:prisms} shows the dramatic difference in appearance that results from this process. After approximately one year of use, visual inspection of the blackened faces showed no apparent degradation.

\section{OPTOMECHANICAL DESIGN}

The optomechanical concept mimics a typical experimental setup in a laboratory, with an optical bench that all optics are mounted to with the optical axis parallel to the bench surface.  The optical bench is then supported by a frame, with panels to seal the instrument from stray light and dust.  The following subsections describe each part in greater detail.

\subsection{Commissioning Optical Bench and Support Structure}
The optical bench is a lightweight, honeycomb core, composite assembly to minimize both weight and changes with temperature.  Manufactured by Vere Inc., the total bench mass is 28.6 kg and is $1.0\times1.15\times0.112\ m$.  The top side of the bench has a grid of M6 threaded holes on a 25 mm spacing.  The bottom of the bench has $3\times3$ equally spaced patterns of four M6 threaded holes in a square pattern, 50 mm on a side.  This enables easy mounting to the support structure when mounted to the telescope, or to the top of a large laboratory optical table during alignment and testing.

\begin{figure}
    \centering
    \includegraphics[width=0.8\textwidth]{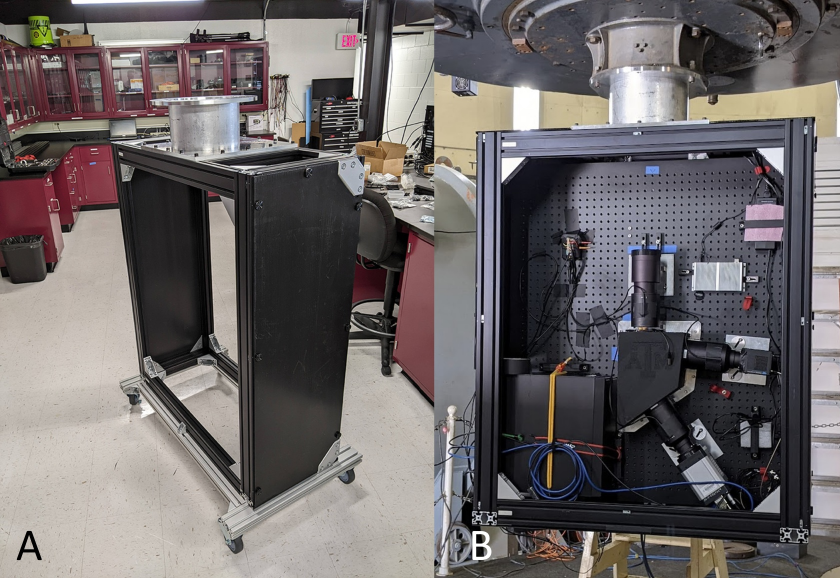}  
    \caption 
    { \label{fig:t-slot-frame}
    \textit{Left, Panel A}: Temporary extruded frame with two of the side panels attached; the composite optical bench has not yet been installed.  A removable cart (silver extrusions) allows ETSI to be rolled around the lab, or into position when mounting to the telescope. \textit{Right, Panel B}: ETSI mounted to the McDonald 2.1 m telescope during preparation for a commissioning run in June 2022. The front cover has been removed.} 
\end{figure}

The optical bench support frame was originally designed to use modular carbon fiber structural tubing, described further below. Supply chain issues and other delays required construction of an alternate support frame (Figure \ref{fig:t-slot-frame}) from extruded Aluminum structural T-slot frame and fittings in order to be ready for the commissioning run allocated telescope time. The optical bench was rigidly supported at three points with flat aluminum plates between the bottom of the optical bench and the T-slot frame.  Removable covers were constructed from corrugated plastic panels. 

This support frame performed admirably during commissioning and initial science observations.  Coincident with initial observations, a refined optomechanical design was developed to incorporate the originally planned carbon fiber structure, while reducing the overall instrument volume.  The temporary extruded aluminum enclosure was limited to targets with Declination less than $+60\degree$ to avoid potential collisions with the South pier of the right ascension axis of the McDonald Observatory 2.1m telescope.

\subsection{Optical Bench and Support Structure}

A lightweight carbon fiber frame, constructed using a modular system from Dragonplate\cite{dragonplate}, was designed.  This system uses preformed carbon fiber tubes and gussets which are bonded together with Scotch-Weld 2216 epoxy and held together during the curing process with pop-rivets.  The tubes are easily cut to length using a standard tile cutting wet saw to minimize carbon fiber dust. More detailed cuts were made using a rotary tool with an abrasive cutting disk while wearing a mask and a HEPA welding fume extractor placed directly above the tool, which captured the majority of the dust. 

\begin{figure}
    \centering
    \includegraphics[width=0.5\textwidth]{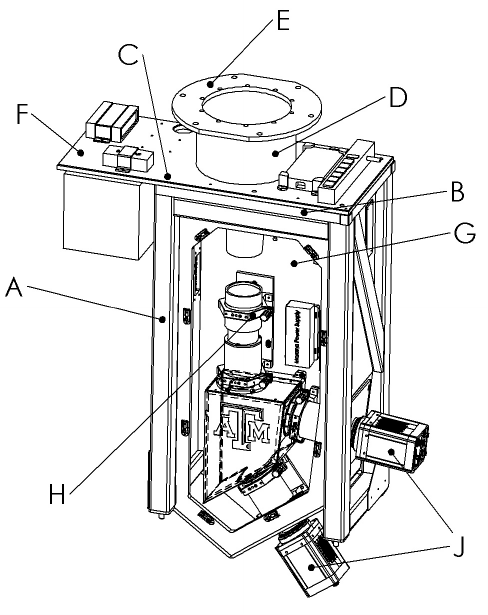}
    \caption 
    { \label{fig:labeled-structure}
    The labeled components are described in detail in the text. The main access panel has been removed in this rendering to show the internal components. [A] Tubular carbon fiber frame, [B] Structural foam and carbon fiber sheet interface plate, [C] Bolts and threaded inserts, [D] Aluminum spacer ring, [E] Telescope interface ring, [F] Aluminum top plate, [G] Structural foam and carbon fiber optical breadboard, [H] Split ring, [J] sCMOS detector systems.}
\end{figure}

Key components are referenced in the text by letter labels in Figure \ref{fig:labeled-structure}. The telescope-facing side of the 50 mm square tubular frame [A] is bonded to an interface plate made from a CNC cut sheet of 25 mm thick structural foam core sheet material with a Divinycell H100 core \cite{divinycellH100} and three layers of plain-weave carbon fiber bonded to each side [B].  The carbon fiber frame and Aluminum plate are bolted together and can be separated for transport.  Both stainless steel bonded nuts and threaded Aluminum inserts epoxied into the ends of the hollow square tubes provide six locations to bolt the two plates together [C].  This dual plate configuration enables a good bonding interface to the carbon fiber frame, while still providing a robust Aluminum interface to the telescope, which is better suited to frequent mounting to the telescope.

ETSI is mounted to the telescope via an Aluminum spacer [D] constructed from a 152.4 mm long, 12.7 mm wall thickness tube and 12.7 mm tool plate telescope interface ring [E] (six 1/2"-13 bolts on a 346 mm bolt circle are used to mount ETSI to the telescope). The other end of the tube bolts to the 12.7mm thick Aluminum front plate [F].

The support structure was assembled in stages, with careful fixturing to ensure alignment of critical pieces.  Because the majority of the structural assembly is held together with Scotch-Weld 2216 epoxy, there are no second chances or repositioning once the epoxy has cured!  The optical breadboard [G] is made of the same 25 mm thick structural foam core carbon fiber sheet, it also makes up one side of the enclosure.  The optics are mounted to the breadboard via stainless steel M6 bonded inserts.  The holes for the inserts are cut via CNC at the same time as the overall profile of the sheet. Standard machining tolerance for the structural carbon fiber sheets is \textpm0.254 mm and \textpm0.127 mm is possible for additional cost.  Optical tolerancing determined the standard \textpm0.254 mm machine tolerances were sufficient for alignment of the collimator, filters, prisms, and cameras and this approach was validated using the commissioning version of the support structure. 

\begin{figure}
    \centering
    \includegraphics[height=8cm]{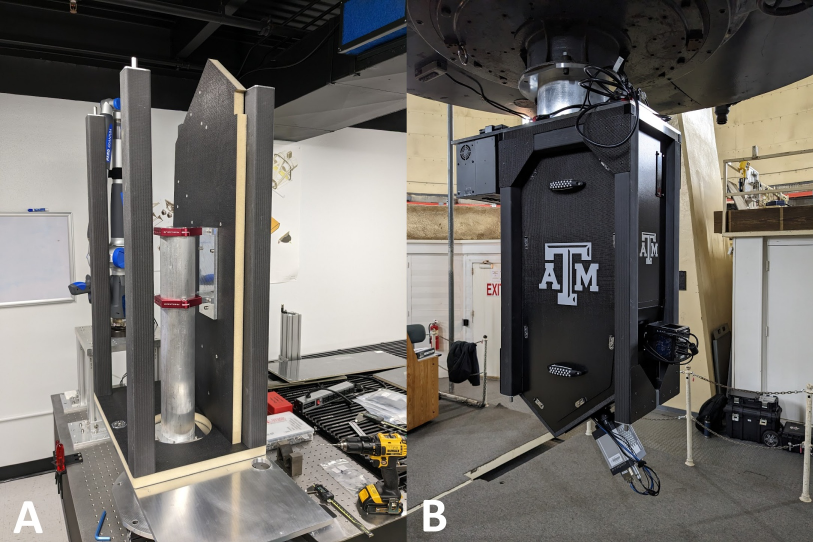}  
    \caption 
    { \label{fig:cf-construction}
    \textit{Left, Panel A}: The carbon fiber support structure was built around an aluminum cylinder mounted to the Aluminum interface plate and shimmed to be perpendicular to the plate.  The collimator optical mount clamps to the cylinder and sets the position of breadboard relative to the optical axis. \textit{Right, Panel B}: The completed carbon fiber structure, mounted to the McDonald 2.1 m telescope.} 
\end{figure}

To ensure the optical bench would be located the correct distance from the optical axis, as well as perpendicular to the telescope mounting interface, an Aluminum cylinder, the same diameter as the collimator lens tube, was machined to length and mounted to the Aluminum interface plate (visible in Figure \ref{fig:cf-construction} (a)).  A coordinate measuring arm (FARO Quantum) was used to measure the perpendicularity and the cylinder was shimmed with plastic shim stock to within 1 arcmin of perpendicular to the telescope interface ring.  The lens assemblies are mounted via two split-ring clamps for each lens tube [H], so the collimator lens tube mount was clamped to the fixture cylinder as well as screwed to the threaded inserts bonded to the carbon fiber breadboard.  The remaining degrees of freedom (rotation and translation about/along the cylinder) were set via measurement with the coordinate measuring arm, referenced to the Aluminum interface plate [F] and telescope mounting flange [E].  Once in place, the split rings were tightened and the cylinder was left in place until the structure had fully cured. Assembly continued in stages, allowing each set of pieces to cure for 24 hours before additional components were bonded.  The epoxy assembly steps took three days to complete. Figure \ref{fig:cf-construction} shows the alignment cylinder and partially-constructed support frame as well as the completed structure mounted to the McDonald 2.1 m telescope.

The remaining side panels are made from 6.35mm thick structural foam core carbon fiber sheets and were also cut to size, including any necessary cut outs, via CNC.  The side opposite the breadboard is mostly open and covered by a 2mm thick sheet of solid carbon fiber, held in place with captured hardware and sealed around the edges with 1mm thick adhesive-backed foam weatherstripping to prevent dust and light intrusion. All edges of the structural foam core sheet were either sealed with epoxy to other surfaces, covered with 1mm thick strips of carbon fiber, or for several curved and difficult to reach places, painted with flat black acrylic paint. Any Aluminum to carbon fiber interfaces were separated with a thin layer of epoxy to prevent galvanic corrosion.

The two detectors [J] extend outside of the enclosure, both to reduce the build up of heat inside the enclosure, as well as provide more flexibility for using alternative detectors. The finished structure is very light weight, the complete instrument when mounted to the telescope is $\sim70\ kg$ and when separated, the aluminum plate and telescope interface and carbon fiber structure (without optics, cameras, etc.) are both easily carried by a single person. The handling cart (not included in the instrument mass) is the heaviest sub-assembly.

\subsection{Optical Mounts}
The collimator and cameras were purchased as complete optomechanical assemblies.  The design of the mounting interface between the lens barrels and supports as well as to the detectors was a cooperative effort with the vendor.  The detectors are mounted with four M3 screws from the camera barrel flange to the body of the detector. The back focal spacing is set by a captured ring, machined to the correct thickness. The focus between channels is matched by placing a back illuminated pinhole at the location of the telescope focal plane.  The pinhole location is adjusted so that the transmitted camera is in focus with a spacer set for the nominal back focal distance.  The reflected channel is then focused to match by adjusting the spacer between the camera barrel and detector with shims.

Each lens barrel is supported by two split rings, mounted to plates that are screwed to the optical bench threaded inserts. The front split ring for each barrel has a hole for a 3mm pin.  The lens barrels have a matching hole for the pin.  The two split rings for each optical tube locate the collimator and cameras on their respective axes, while the pin between each tube and the first split ring set the axial position as well as clocking of the cameras.  The lens barrels can be removed for transport to the observatory and repeatably mounted upon arrival at the telescope.  ETSI has been assembled/disassembled at McDonald Observatory several times during commissioning and science observations, each time the optics and cameras have been mounted to the breadboard with no additional alignment in only a few minutes. The detectors remain mounted to the lens barrels for transport.

The position of all lens barrels relative to the multi-chroic and prisms are set by the locations of the mounting screws and threaded inserts in the carbon fiber breadboard.  Each optic mount uses three screws to set the location, a shoulder screw in a close fitting hole, a shoulder screw in a close fitting slot and a third screw in an oversized hole.  In practice, the alignment tolerances between collimator, multi-chroic, prism, and camera barrel are loose enough that no further adjustment has been necessary after attaching all optical mounts to the breadboard.  Alignment was checked using a coordinate measuring arm and each lens barrel was within $\sim8$ arcmin of the optical axis. This level of misalignment was undetectable in lab images and negligible when modeled with Zemax. The self referencing technique used in the ETSI instrument removes most systematic errors that would typically arise due to misalignment or flexure, thus eliminating the need for extremely tight instrument tolerances.  The $f/13.7$ focal ratio of the McDonald 2.1 m telescope also eases the alignment tolerances.

\begin{figure}
    \centering
    \includegraphics[width=0.8\textwidth]{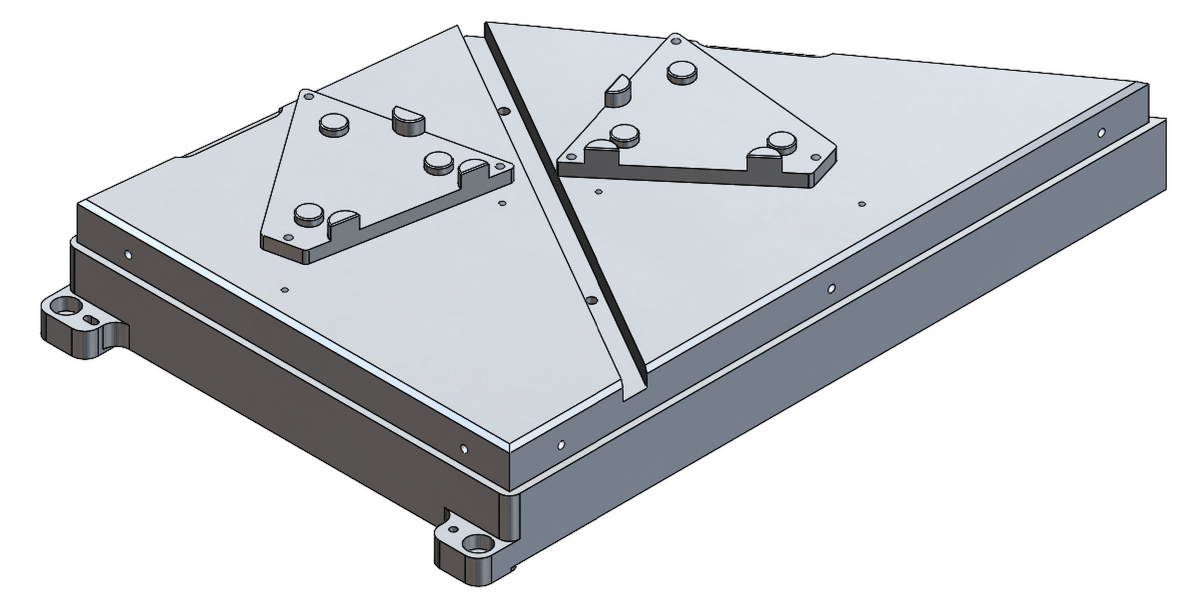}
    \caption{\small{The prism base.  A slot for the multi-chroic mount is visible with machined prism-locating points on either side.  The prism base was machined out of a single piece of 63.5 mm thick tool plate, the underside has been light-weighted to a 5mm wall thickness.}}\label{fig:prism-base} 
\end{figure}

The multi-chroic and prisms are mounted to an Aluminum support plate that was machined out of a single piece of tool plate (Figure \ref{fig:prism-base}).  It has a slot to hold a plate at the $30\degree$ multi-chroic angle.  The multi-chroic is mounted into this plate in a circular recess and held in place with a retaining ring with o-ring spacer.

The prisms are mounted in a 3-2-1-style fixture consisting of three pads machined into the mounting plate to set the prism height and rotation along two axes, two cylindrical contact points on one face of the prism and one cylindrical contact point on a second face.  An Aluminum plate clamps the prisms onto the prism base via three long screws, one at each corner.  A thin textured rubber sheet is placed between the top of the prism and the clamping plate to allow for differential expansion and contraction due to temperature changes. Baffles were 3D-printed and internal surfaces coated with adhesive-backed light absorbing Fineshut SP urethane foam (See \cite{Schmidt2022} for total reflectivity measurements of this material. Average reflectivity over the ETSI bandpass is $<1\%$).

The prism and multi-chroic mount assembly is covered with a 3D-printed cover.  The entire assembly can be removed from the breadboard by removing three M6 screws.

\section{ELECTRONICS \& SOFTWARE}
\subsection{Detectors}
The recent availability of high quality optical CMOS sensors\cite{Andor, Teledyne} (often referred to as sCMOS or scientific CMOS) has proven crucial to the success of ETSI. An early ETSI prototype used a commercially-available CCD.  There were no issues with the data quality, however even at the highest read out speed the observing efficiency was very low.  In order to measure the small color changes during a transit, averaging over many measurements is required and the overall sensitivity is driven by the total number of photons detected.  For bright targets especially, the exposure time can be shorter than the read out time of the CCD, resulting in many missed photons.  Additionally, read noise becomes an issue at all but the slowest read out rates.

One possible solution considered was to switch to a frame transfer Electron Multiplying CCD (EMCCD). Frame transfer allows for faster read rates and the electron multiplication process can essentially eliminate read noise under certain observational conditions.  However, for any pixels where the shot noise is above the readout noise, the SNR is reduced by a factor of $\sqrt{2}$ due to the stochastic nature of the electron multiplication process. This means that EMCCD's are best suited for low photon counts.  The electron multiplication process also effectively reduces the possible dynamic range, so a 16-bit EMCCD has an equivalent, or even lower dynamic range than the 12-bit readout (many sCMOS sensors have a dual gain, 12-bit ADC) of an sCMOS detector for higher EMCCD gain values.

The latest generation of sCMOS devices solve essentially all of these issues, with future generations promising even better performance.  Our chosen detector, the Andor Marana sCMOS camera is based around the GSENSE400BSI sensor, a back-illuminated, $11\mu m$ pixel, $2048\times2048\ px $ array.  The large pixel size and high quantum efficiency (peak QE of 95\%) make for efficient collection of photons and the readout time of $\sim20\ ms$ enables a minimum of observation time lost to readout. Read noise is $<2e^{-}$ at all frame rates. One aspect of the current generation of sensors we are still evaluating is the inclusion of a dual amplifier architecture, where each pixel is read by two amplifiers, each with a different gain.  Two readout modes are available, either a ``high gain" mode, in which a single amplifier is used and the output is a 12-bit image, or a ``high dynamic range" mode where, on a pixel-by-pixel basis, a decision is made on which amplifier to use (if the high gain amplifier is saturated, the low gain amplifier is used).  The two data streams (high and low gain) are digitized at 12-bits and are then set to a common linear scale and combined to a single 16-bit image. Due to the rapidly evolving sCMOS camera availability, our reflected camera has changed several times as we evaluate different models.  Most reflected channel observations have been made with a Teledyne Kinetix sCMOS camera.  The different format (6.5 micron pixel, 3200 x 3200 px array) has essentially the same field of view as the Marana camera on the transmitted channel.  A Teledyne Kuro sCMOS camera is the current reflected channel camera (11 micron pixel, 2048 x 2048 px array) and uses the same GSENSE400BSI sensor as the transmitted channel Andor Marana. A recent award (From the Mt. Cuba Astronomical Foundation) has enabled the purchase of matching Teledyne COSMOS-10 sCMOS cameras, which have 10 micron pixels in a 3260 x 3260 px array and most importantly 16-bit digitization from a single readout. This will result in 0.164"/px and a potential 8.9' field of view. However, the usable field of view will be less as the original optimization was for a smaller field of view so a combination of image quality reduction and vignetting will limit the  field to approximately 7 arcminutes on the McDonald 2.1 m telescope.  We expect to integrate the new cameras soon after they are delivered in early 2025.

Currently we are collecting any transit data in the ``high gain", 12-bit output mode.  This can result in higher data rates ($\sim10$ frames per second for an 8th magnitude star on the McDonald 2.1 m telescope) to avoid saturation, but was decided to be the conservative approach until a full analysis can be performed on the dual-readout 16-bit data.  Over a full transit on bright targets, the overall data volume is substantial.  For individual exposure times less than a second, the data is saved as co-added images ($\sim10-15$ frames per co-add) to reduce the required disk space while still providing sufficient time resolution. 

\subsection{Shutter, Temperature Monitoring, \& Communication}
\begin{figure}
    \centering
    \includegraphics[width=8cm]{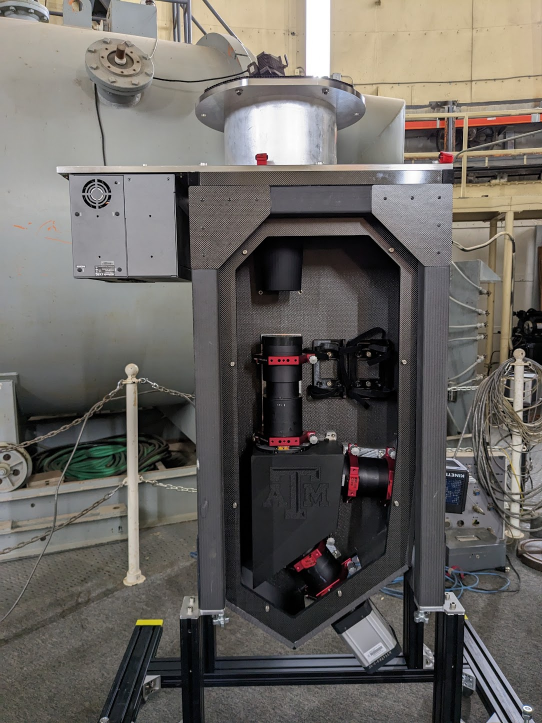}
    \caption{\small{ETSI during assembly at McDonald Observatory. The red split rings holding the lens barrels are visible. The multi-chroic and prisms are mounted inside the black 3D-printed cover with the Texas A\&M Logo.  The black extruded aluminum frame holding ETSI can be rolled around the observing floor and is removed once ETSI is mounted.  Spring-loaded casters allow for any misalignment as ETSI is lifted up to the back of the telescope for mounting. }}\label{fig:assembly-dome} 
\end{figure}

ETSI has essentially no moving parts with the exception of a Uniblitz NS65B bi-stable optical shutter controlled with a Uniblitz VED24 shutter driver.  The shutter is located between the collimator and multi-chroic to allow for acquisition of dark and bias calibration frames and is mounted to the entrance port of the 3D-printed prism and multi-chroic mount cover.  It is not used during normal observations due to the high frame rates (the rated continuous operation frequency is 1Hz). Instrument internal and external temperatures are monitored via low-cost 1-wire DS18B20 temperature probes.  All control and interface signals are via USB.  Two operational modes have been used.  The original control interface was a FireNEX-5000Plus USB 3.1 fiber optic extender, which has a four port hub on the remote unit and is then connected to a local unit via duplex single mode fiber. This allows the control computer to be located in the control room.  In practice even an armored casing fiber was not robust enough for long term use as the control cable drapes from the back of the instrument, across the observatory floor and to the control room making consistent strain relief challenging. In the current configuration an industrial control computer is mounted to the Aluminum top plate that is the interface to the carbon fiber enclosure (instrument top left in Figure \ref{fig:assembly-dome}). Instrument control is accomplished via remote desktop software.  The only external cabling between ETSI and the observatory is 120V power and Ethernet.

\subsection{Camera Control}
The sCMOS cameras are controlled by a Python program.  The Andor Marana camera interfaces with the Andor SDK and andor3 Python interface\cite{Tapping2021}. The andor3 Python interface includes utilities that control the frame buffer and frame delivery to the rest of the program. The Teledyne Kinetix sCMOS camera interfaces with Teledyne PVC and PyVCAM\cite{Hanak2024}. The Teledyne Kuro sCMOS camera interfaces with Teledyne PICam\cite{Liakat2024}. The Teledyne interfaces were edited to use similar frame buffer and frame delivery systems as the andor3 Python interface. A user interface allows observers to set up each acquisition sequence. Settings such as exposure time, co-add count, and gain (12-bit or 16-bit) are selected by the user and sent to the camera.  When the camera is commanded to begin taking frames, the frame buffer waits for each new frame and delivers the frame to a callback function. The callback function saves each frame as a FITS file with informational headers that include position, environmental, and tracking information from the McDonald Observatory 2.1 m telescope control system.  The transmitted camera sends a TTL trigger pulse to the reflected camera to ensure image acquisition is synchronized. Both cameras have native trigger I/O interfaces accessed via multi-pin to coax breakout cables.  The breakout cables for trigger out/in are directly connected to each other. The transmitted camera initiates the exposure sequence and the reflected camera starts its exposure on the rising edge of the transmitted cameras 'start exposure' TTL pulse.
\subsection{Interface with McDonald Observatory 2.1 m telescope TCS}
The position and tracking information from the McDonald Observatory 2.1 m telescope control system (TCS) is collected by connecting to the TCS network and requesting the information once every second. Between each request, the information is stored so it can be inserted into each frame’s FITS header as each frame arrives. Additionally, a frame or stack of frames is sent to the 2.1 m telescope's auto-guiding software, Guide82, every 10 seconds. Guide82 also reports estimated Full Width at Half Maximum (FWHM) of the images and includes an automated focusing routine.

\section{THE ETSI DATA REDUCTION PIPELINE}
\label{sec:data-reduction}
During commissioning a Quick Look Pipeline (QLP)\cite{Schmidt2022} was used to provide real-time analysis of target photometry concurrent with ETSI observations while users are at the telescope. Real-time estimates of the white light curve of the target star helped confirm that ETSI was functioning as expected. The QLP is no longer used for regular science observations. 

The full data reduction suite for ETSI consists of three main components: image pre-processing, photometry extraction, and systematics removal. ETSI science images meant for relative color measurements are not calibrated using traditional photometric methods. We found that the sCMOS systems are remarkably stable and that standard flat fielding, bias subtraction, and dark correction was unnecessary to meet our science goals. These steps added noise to the data and we were unable to generate a master bias and flat frames with noise values below the photon precision we hoped to achieve.  Instead, images are co-added to a desired timescale (typically 1 minute) in order to reduce file size and increase measurement precision.

Photometry is extracted from each image typically using fixed-size aperture photometry (though both adaptive aperture photometry and PSF fitting are available). A master image is generated to locate the position of each spectral PSF. We found traditional star finding routines failed with the ETSI PSF and did not order each spectral PSF in a convenient fashion. Instead, the initial location of a single PSF of the target star (and comparison star, if necessary) is manually selected on the master frame to provide an initial solution for the location of the star. The ETSI pipeline then automatically identifies the position of each PSF using the known dispersion of the instrument and centroids these positions to find the center of mass of each PSF within a box of half the size of the elliptical apertures. The ETSI PSF is then extracted from each image (after re-centroiding for small shifts between images) using the \texttt{PYTHON} photometric software \texttt{photutils} \cite{photutils}. The stellar flux is extracted using a fixed-size elliptical aperture of $40\times25$ pixels for the transmitted data and a fixed-size elliptical aperture of $65\times30$ pixels for the reflected data. The difference in aperture sizes reflect the difference in pixel scale, which leads to more pixels per PSF on the reflected camera. 

The sky background was estimated using a sky aperture of identical size to the target aperture which was placed above and below the PSF of each bandpass, typically between $150-300$~pixels of the target aperture. The sky apertures were moved closer to the target aperture or farther from the target aperture in order to avoid flux contamination from nearby stars. The fluxes in each sky aperture were summed, the two measurements were averaged, and the resulting value was subtracted from the total summed flux of the target aperture. The target flux was then converted to e$^{-}$/s using the total exposure time of the co-add (typically 60~s) and the gain of the detectors (0.61~e$^-$/ADU). The light curves were then converted to an instrumental magnitude with the standard formula

\begin{equation}
    m_{i} = 25 - 2.5\log_{10}(f_i)
\end{equation}

Common-path systematics (such as those from airmass, cloud cover, and atmospheric color-terms) were removed from each light curve using a time-averaged ``trend" light curve unique to each bandpass. The following process is completed for exoplanet targets. These trend light curves were generated by linearly combining all other available spectral bandpasses (across both cameras) with the formula,

\begin{equation}
    t_{i} = \sum_{j=0}^{N} (c_j \cdot m_j + b_j) \text{; where } i \neq j
\end{equation}

where $t_i$ is the magnitude of the trend of the $i^{th}$ spectral bandpasses, $N$ is the number of bandpasses, $m_j$ is the magnitude of the $j^{th}$ bandpass, $c_j$ is the best-fit scaling correction for the $j^{th}$ bandpass, and $b_j$ is the best-fit shift for the $j^{th}$ bandpass. This trend is generated using the \texttt{LinearRegression} function in the \texttt{skearn.linear\_model PYTHON} library.

The final normalized light curve for each bandpass was calculated by subtracting the trend light curve from the target light curve. This type of spectral band referencing is a common method to reduce systematics in transmission light curves and is a proven technique for exoplanet light curve analysis to reduce systematic noise sources \cite{Cartier:2017, Louden:2017, Stevenson:2020, Kirk:2021, Ahrer:2022}.  

We found this method to vastly improve the capabilities of removing systematics from the target light curves over more traditional methods, such as comparison star referencing. In comparison star referencing, the light curve of a nearby bright star is subtracted from the light curve of a target star. Typically this is a useful way to remove systematics because both stars were simultaneously observed through roughly the same atmosphere and appear at nearby positions on the detector. However, because the light from the two stars did not travel through exactly the same path prior to landing on the detector, and because not every star in the sky has a nearby, similar magnitude companion, the comparison star method is not a fool-proof method to remove systematics.

\begin{figure}
    \centering
    \includegraphics[width=\textwidth,height=\textheight,keepaspectratio]{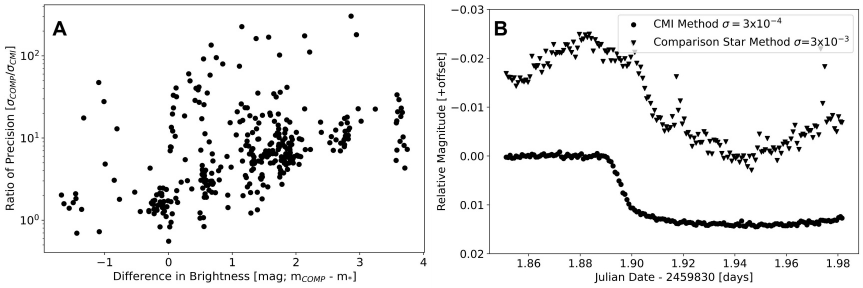}
    \caption{\textit{Left, Panel A}: A comparison of the difference in brightness between the target star and comparison star and the ratio of the out-of-transit precision measured for 330 light curves observed on 22 separate nights using both the comparison star method ($\sigma_{COMP}$) and the CMI method ($\sigma_{CMI}$). On average, we find the CMI method improves light curve precision by a factor of $\sim1.5$ for comparison stars within 1 magnitude of the target star or brighter. We only found the comparison star method to improve the precision on 7 out of 330 light curves. \textit{Right, Panel B}: The achieved dispersion in the light curve of WASP-33~b observed on September 08, 2022 when using the CMI method (black; $\sigma=3\times10^{-4}$) and when using a traditional comparison star method (red; $\sigma=3\times10^{-3}$). We find a factor of 10 improvement in out-of-transit precision quality when using the CMI method for this light curve.}\label{fig:precision}
\end{figure}

We compared the precision achieved by both the CMI method and the comparison star method for 330 light curves observed on 22 separate nights. We found, on average, the CMI method achieved better precision for ETSI light curves by a factor $\sim1.5$ when the comparison star is within 1 magnitude of the target star (or brighter) and we only found the comparison star method to provide better precision for 7 out of 330 light curves. For example, we found the dispersion in the transit light curve of WASP-33~b dropped from $\sigma=3\times10^{-3}$ to $\sigma=3\times10^{-4}$ and most of the inherent stellar variability was removed from the transit light curve when using the CMI method as compared to the comparison star method as shown in the right panel of Figure~\ref{fig:precision}. The distribution of the ratio of the achieved precision of both methods for all 330 light curves is shown in the left panel of Figure~\ref{fig:precision}.

While this method is effective at removing non-astrophysical signals, it also removes common, non-color-dependent astrophysical signals, such as the mean transit signal (white-light transit), because this signal is common to all bandpasses. Therefore, a model of the white-light transit signal was injected into each bandpass's de-trended light curve prior to measuring the transit depth at each wavelength. 

We found injecting the white-light curve transit was useful for two reasons. First, we found transit fitting routines were more capable of measuring realistic depths after a model transit was injected than they were at measuring residuals. This was particularly important for residuals containing a ``negative" transit depth because the transit was shallower than the white-light signal. Second, because we injected the same white-light transit signal into each bandpass we knew the ground truth mean transit depth \textit{a priori}. This allowed for a much more convenient comparison of the recovered signal to a flat-line, which could indicate a null-detection.

\begin{figure}
    \centering
    \includegraphics[width=0.8\textwidth]{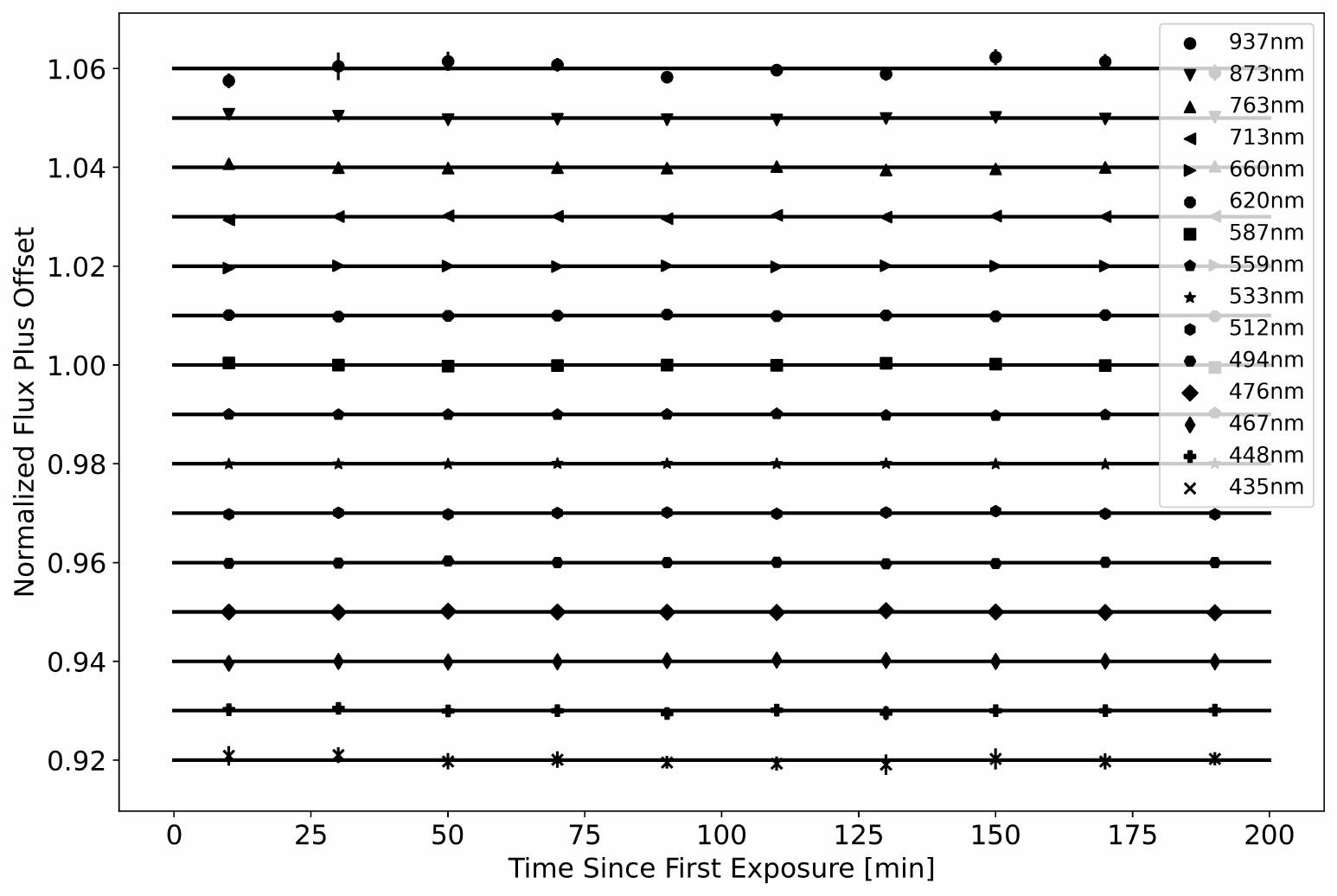}
    \caption{\small{Light curves for the 15 ETSI bandpasses for the A0 star HD~9711 over the course of $\sim200$ minutes. Each bandpass has been cleaned using a linear combination of the other 14 bandpasses. The data has been binned into 20 minute intervals and offset by $1\%$ for clarity. The solid lines are not fits to the data, but are simply a line with 0 slope at each offset. The error bars are plotted on the figure and are roughly the size of each data point or smaller. The standard error of the mean of these light curves suggest, on average, the observations could have identified a color change of $0.006\%$.}}\label{fig:hd9711} 
\end{figure}

Next, the precision of the ETSI instrument was investigated using the light curves of the A0 star HD~9711 (V=9.98)\cite{2000A&A...355L..27H} in order to identify the minimum level of detectable color change ETSI could achieve over 3 hours of observation, a similar timescale to an exoplanet transit. A-type stars are expected to have little-to-no changes in their color, and by investigating the dispersion of ETSI color curves for HD~9711 a limit can be placed on the minimum detectable color change. The change in color for each bandpass was found to be consistent with a zero slope line in all cases (the coefficient of determination, $R^2$ was $\sim0$). The standard error on the mean (binned in an interval of 200 minutes) was found to be $0.006\%$ as shown in Figure~\ref{fig:hd9711}. This color change is consistent with the systematic error achieved by the Hubble Space Telescope (HST) of $\sim0.001-0.01\%$ on similar timescales \cite{Sing:2016}. This means ETSI can provide HST-quality precision for the measurement of color changes over the timescale of an exoplanet transit.

Finally, we expect future iterations of the pipeline to have the capability to substitute its self-referential photometry functions with more traditional methods to remove detector systematics from the ETSI photometry. These methods could include:
\begin{itemize}
    \item Comparison star detrending: Typically, two stars of similar magnitude should not vary in identical astrophysical patterns between images, even if they were the same spectral type or variable type. In many cases, the bandpass-specific flux from a comparison star can be binned using a moving mean and used to normalize the target star's bandpass flux and remove common systematics \cite{Wang2013, Vanderburg:2014, Oelkers:2018ffi}.   
    \item Observational Bias detrending: Ground-based observations will suffer from changes due to airmass, extinction from clouds, changes in the temperature of the detector, and the slight changes of the position of the star on the detector. A trend could be generated through a linear or polynomial combination of these factors to remove systematics which are not found to be correlated between the target star and its comparison.
\end{itemize}

\section{SCIENCE WITH THE ETSI INSTRUMENT}
ETSI was primarily designed to carry out exoplanet characterization and vetting via spectral transmission measurements at $R=\lambda / \delta\lambda \approx20$.  Below we describe this and the other potential scientific capabilities of ETSI in more detail. 

ETSI characterizes exoplanet atmospheres by producing spectra in the visible ($430<\lambda<975$~nm) in 15 spectral bands. This is comparable to the spectral resolutions reported by other transmission spectroscopy measurements, which are typically collected at higher resolutions but binned down to increase the signal-to-noise ratio and study the same spectral features and similar bandpasses as ETSI \cite{2008MNRAS.385..109P, 2015ApJ...814...66K} or only focus on a very narrow spectral range at high resolution \cite{2015A&A...577A..62W, 2020MNRAS.493.2215G}. The location of the ETSI spectral band filters are designed to coincide with spectral features of interest in exoplanet atmospheres. Specifically, ETSI transmission spectra measurements are sensitive to the presence of atoms (potassium and sodium), molecules (methane, water, and TiO), and clouds and aerosols in the exoplanet's atmosphere (Rayleigh scattering). ETSI bandpasses were specifically tuned for characterization of transiting gas-giant planets, but additional filters could be produced to match other targets.

These self-referenced differential measurements serve as a proxy for the measured atomic and molecular features in the exoplanet's atmosphere. These features can be modeled and the atmospheric composition of the planetary atmosphere can be inferred. Spectral band referencing is used to reduce systematics in the transmission light curves and is a proven technique for exoplanet light curve analysis to reduce systematic noise sources \cite{Stevenson:2020,Ahrer:2022}. Residual systematics can be removed with typical detrending techniques (i.e. polynomial detrending or Gaussian process regression \cite{Cartier:2017,Louden:2017,Kirk:2021}) in combination with time-averaged light curves from reference stars in the field of view. The best-fit exoplanet atmospheric model can then be found by using standard exoplanet atmospheric retrieval tool kits, specifically the open source \texttt{python} code \texttt{petitRADTRANS} \cite{Molliere:2019,Molliere:2020} in combination with the \texttt{emcee} Markov Chain Monte Carlo framework \cite{Macky:2013}. Results of the first sample of exoplanet transits observed with ETSI are reported in Oelkers 2025 \cite{2025AJ....169..134O}.

Transits rarely required a complete night for observation and a significant number of night time hours were available for other observations while waiting for the transit observation window.  This time was used to explore other, non-transit observations that could take advantage of the simultaneous multi-band capabilities of ETSI. 

\begin{figure}
    \centering
    \includegraphics[width=0.8\textwidth]{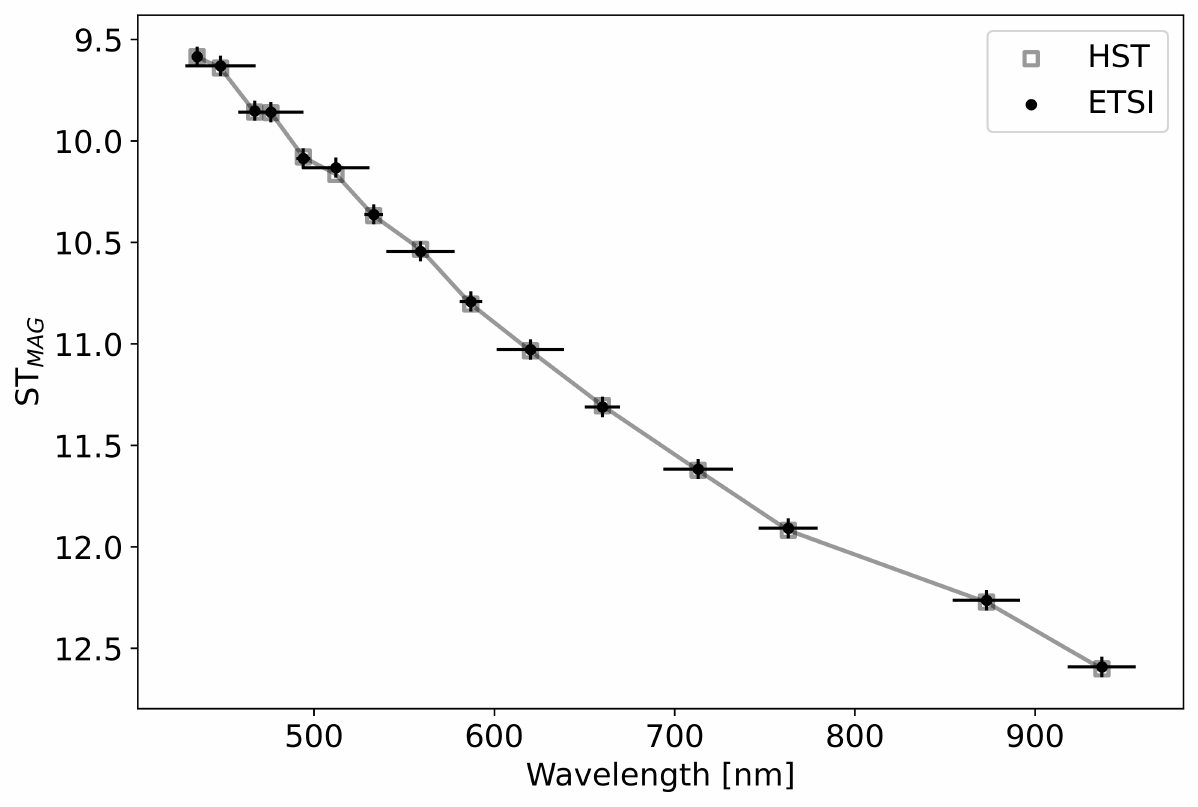}
    \caption{\small{Comparison of ETSI spectrophotometric measurements with a HST standard star.\hspace{3cm} }}\label{fig:bd28} 
\end{figure}

Observations of existing spectrophotometric standard stars with ETSI can be used to expand the number and sky coverage of spectrophotometric standard stars.  The common path and simultaneous nature of ETSI observations reduces the influence of many common photometric errors and dramatically improves the observing efficiency as there is no observing time lost to filter changes or re-observation due to weather changes half way through a filter sequence. An example is shown in Figure \ref{fig:bd28}.

ETSI also excels at observing a wide variety of variable sources. Simultaneous measurement of 15 filter bands combined with the low read noise, high cadence imaging modes enabled by sCMOS detectors can be used to monitor sources that are variable on a wide range of timescales.  Our highest cadence observations to date are 10 frames per second, sustained for 3-6 hrs. This cadence was required to avoid detector saturation, so the data was typically saved as co-added 1.5 second equivalent exposures to reduce data storage requirements, but that is not required if the higher time resolution is desired. We have observed sources that are variable on a variety of timescales including white dwarfs (minutes), exoplanet transits, RR Lyrae (hours-days), and several supernovae (days-weeks). Representative observations of a variety of targets are shown in Section \ref{sec:preliminary}.

\section{COMMISSIONING}
ETSI achieved first light April 19-24, 2022 on the McDonald Observatory 2.1 m telescope with only the transmitted channel, allowing simultaneous photometry of eight spectral bands.  Observations with both transmitted and reflected channels (15 bands) began June 6-21, 2022 and the final carbon fiber support structure was first used April 10-14, 2023, all on the McDonald Observatory 2.1 m telescope.

\subsection{Preliminary Results}
\label{sec:preliminary}
\begin{figure}
    \centering
    \includegraphics[width=0.8\textwidth,keepaspectratio]{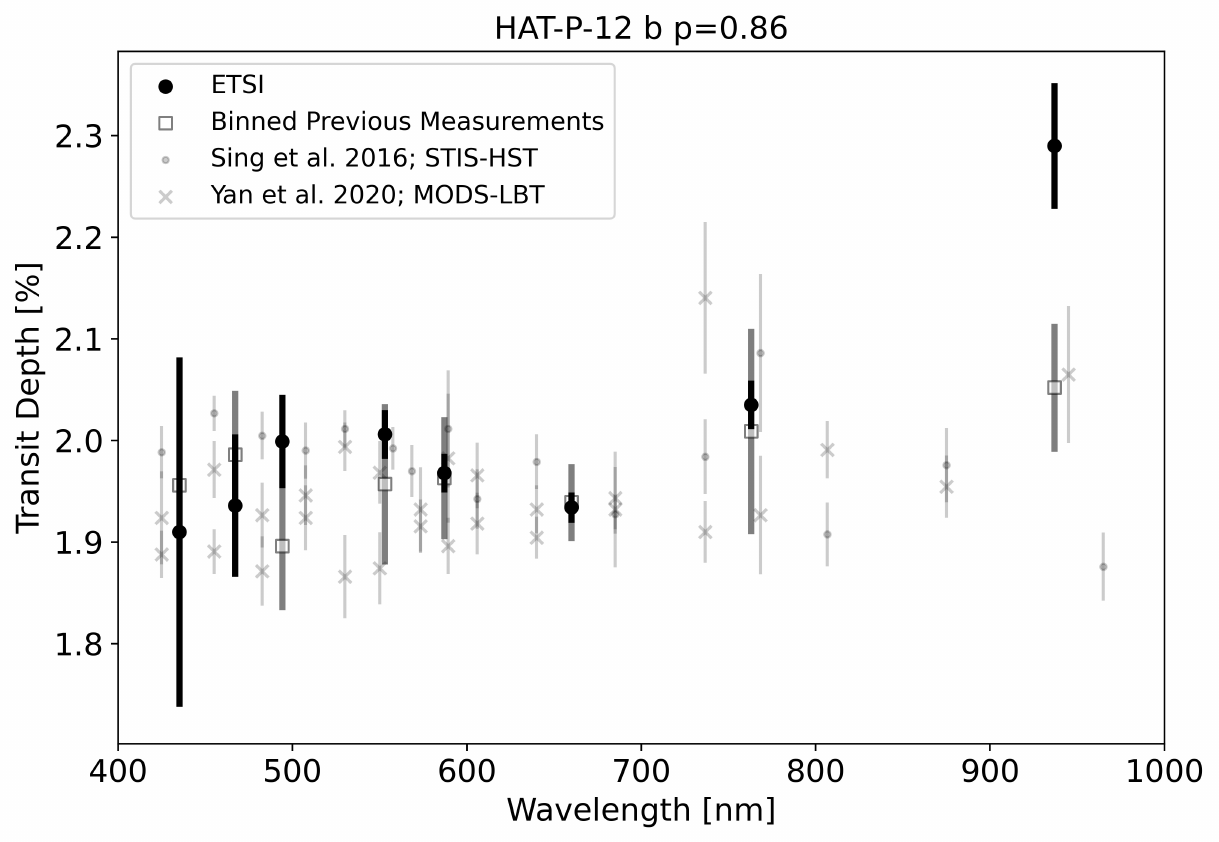}
    \caption{\small{A comparison of the measured transit depths for HAT-P-12b using data measured with HST  \cite[grey circles(original data)]{Sing:2016}, LBT \cite[grey crosses (original data)]{Yan:2020}, and ETSI (black dots). The grey square represent the HST and LBT data binned to match the ETSI bandpasses. The uncertainties in the bins represent the scatter in the original data's spectrum at the given instrument wavelengths. The measurements are consistent within the uncertainties and an Anderson-Darling test returned a p-value $p=0.86$, suggesting the measurements from HST, LBT, and the ETSI instrument are consistent.}}\label{fig:hat12p} 
\end{figure}

Light curves have been investigated for several exoplanets both during and after instrument commissioning.  An in-depth discussion of these observations is forthcoming in Oelkers et al.\cite{2025AJ....169..134O}. One such object was HAT-P-12~b (HAT-P-12, V=12.84 \cite{2009ApJ...706..785H}) which was targeted because it has been observed with multiple other observatories and instruments, including the Hubble Space Telescope\cite{Sing:2016}, and can provide a comparison with previous results. A single partial (82\% of the transit duration) transit was observed with ETSI during an observing run in June of 2022 (details of the observation and analysis are in Oelkers et al.\cite{2025AJ....169..134O}). The transit depths measured using ETSI in combination with a 2-m class telescope (that costs $\sim\$160$/night) were then modeled and statistically compared with the transit depths measured using HST (that costs $\sim\$100,000$/hour) \cite{Sing:2016} using a two-sample Anderson Darling test. The p-value of the test was found to be $p=0.86$ indicating the measurements from HST and ETSI are consistent for both targets (See Fig. \ref{fig:hat12p}). The achieved precision of ETSI, lack of shown color change in an A0 star, and the recovery of consistent transit depths between ETSI and HST provides substantial proof of concept evidence that an atmospheric detection can be made using a 2~m class ground-based telescope and ETSI.

Exoplanet atmosphere characterization using ETSI could be leveraged to prioritize the most interesting targets for follow up observations using JWST. Over-subscription rates for JWST are increasing (7.3x for cycle 2, \cite{JWST:sub23} 9x in cycle 3 \cite{JWST:sub24}.) so any pre-filtering of potential targets that can occur will be a benefit to the overall community.

Several spectrophotometric standard stars were observed with ETSI to estimate the level of precision it could achieve for flux calibration in the following manner. First, we zeropointed ETSI observations of the standard star HZ44 to the CALSPEC library of composite stellar spectra from HST \cite{Oke:1990, CALSPEC, Bohlin:2020, Bohlin:2022}. We then applied these zeropoints to the ETSI observations of the standard star BD+28-4211 and compared our calibrated flux measurements to the CALSPEC composite stellar spectra of the star.  As shown in Figure \ref{fig:bd28}, ETSI produces excellent spectrophotometric calibration across all 15 filter bands (within 0.9\%) for stars with brightness 5 $< V <$ 14 magnitudes with less than 1 hour of observations.

\begin{figure}
    \centering
    \includegraphics[width=0.8\textwidth]{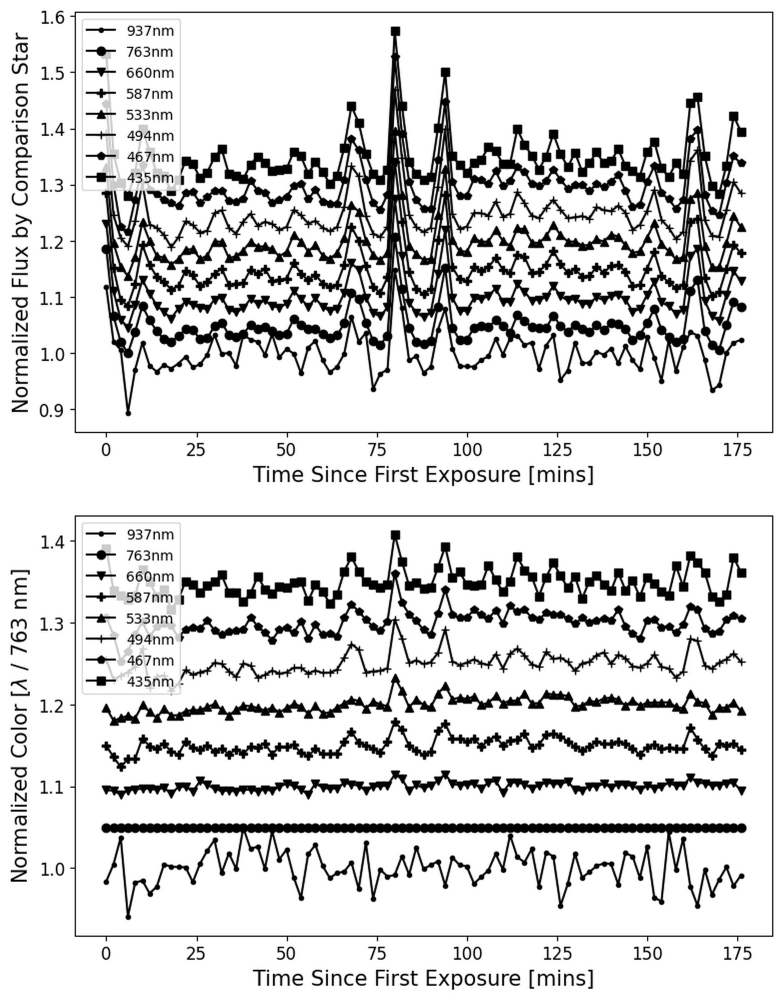}
    \caption{\small{Light curve (top) and simultaneously-measured optical colors (bottom) for the $\sim$13.6 magnitude pulsating white dwarf GD358. Note easily detected $\sim$ 0.05 magnitude bluer color of the $\sim$30\% brightness increases at around 80 and 90 minutes. Data offset by 0.05 magnitudes for clarity.}}\label{fig:gd358} 
\end{figure}

Exoplanet transit measurements look for color changes on hours-long timescales.  ETSI also excels at measuring color variability on much shorter time intervals. The white light variability of the white dwarf GD358 has been previously measured\cite{Winget:1982}. ETSI observations reveal possible color variability on similar timescales (5 s reported in the literature\cite{Winget:1982} and 20 s measured with ETSI).  Normalixed flux and normalized color plots of GD358 are shown in Figure \ref{fig:gd358}.

\begin{figure}
    \centering
    \includegraphics[width=0.7\textwidth]{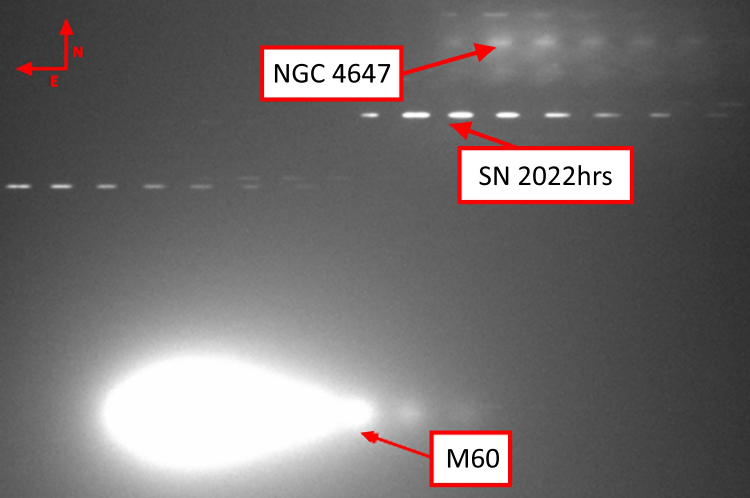}
    \caption{\small{ETSI multi-band image of SN2022hrs. We have achieved $<1$\% photometric precision measuring supernova despite the complicated background from nearby galaxies. The multi-band nature of ETSI images could enable both supernova type and redshift identification using a single ETSI exposure.}}\label{fig:sn2022hrs} 
\end{figure}

As demonstrated by the observations of HZ44 and BD+28-4211, ETSI observations can reach 1\% spectrophotometric precision across 15 filter bands with only a short ($<1$\ hr) observation time. Sensitivity to color changes, as demonstrated by the observations of the A0 star HD 9711, make ETSI a useful tool to rapidly categorize the type and redshift of bright supernovae.   We are exploring the brightness and precision limits of these measurements as well as developing additional data reduction techniques to deal with the complicated background introduced by the host galaxy and other nearby galaxies. We currently use the same multi-band filter that was optimized for exoplanet transit spectroscopy, but are exploring other bandpass combinations that are better suited to supernova characterization. An example frame from an observation of SN2022hrs is shown in Figure \ref{fig:sn2022hrs}. Expected alert rates from LSST are  $\sim$200 SNe alerts per visit\cite{Graham2024}. Rapid follow-up observations with other telescopes will be critical to extracting useful information from these alerts.

\section{CONCLUSIONS}
ETSI is a unique instrument that leverages the advantages of simultaneous multi-band photometry and is made possible by several key technologies, including multi-band interference filters and sCMOS detectors. Early results are promising and indicate ETSI will be a highly productive instrument, characterizing exoplanet atmospheres in only a few transits and may be useful for JWST target prioritization and rapid follow-up of LSST transient alerts. We are actively exploring ETSI's applicability to other science targets such as supernovae, variable stars, and rapid characterization of other transient events. ETSI has observed more than 90 nights between June 2022 and June of 2024 and has obtained transmission spectra of 21 exoplanets \cite{2025AJ....169..134O} during these observations as well as observations of white dwarfs, brown dwarfs, standard stars, planetary nebulae, a stellar occultation of Titan, supernovae, and blazars.  Work is ongoing to refine data reduction techniques for the unique multi-band images generated by ETSI.

\clearpage
\subsection* {Disclosures}
The authors have no conflict of interest to declare.

\subsection* {Code, Data, and Materials Availability} 
The data that support the findings of this article are not publicly available as they are included only to demonstrate the range of scientific capabilities of ETSI. They can be requested from the author at \linkable{lschmidt@yerkesobservatory.org}

\subsection* {Acknowledgments}
Texas A\&M University thanks Charles R. '62 and Judith G. Munnerlyn, George P. '40 and Cynthia Woods Mitchell, and their families for support of astronomical instrumentation activities in the Department of Physics and Astronomy. The ETSI project is funded by the NSF MRI grant no. 1920312 and the Mt. Cuba Astronomical Foundation.  The authors would like to thank John Kuehne for his assistance with integrating ETSI control software with the 2.1 m TCS and Coyne Gibson for his support in developing the mechanical interface between ETSI and the 2.1 m telescope.


\bibliography{report}   
\bibliographystyle{spiejour}   

\appendix
\clearpage
\section{Optical Prescription}
\begin{figure}[h]
    \centering
    \includegraphics[width=0.95\textwidth]{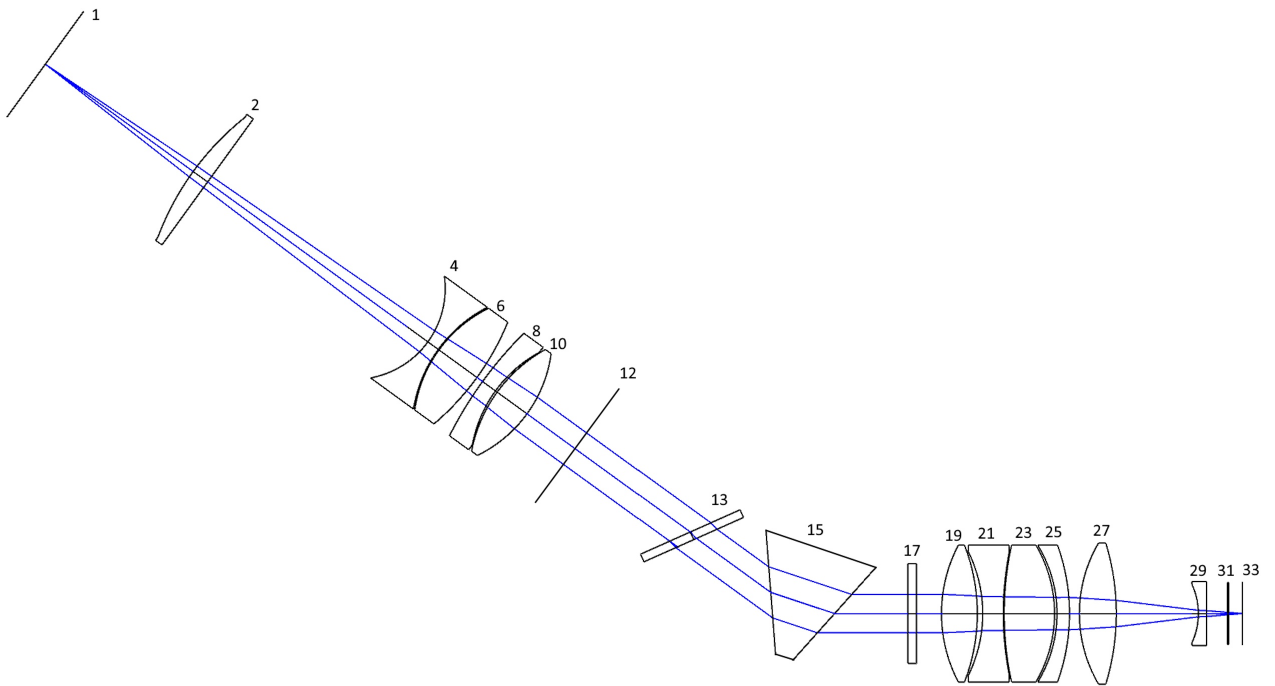}
    \caption{\small{ETSI optical layout, showing only the transmitted channel. Each element is identified by the number of their first surface in Table }\ref{tab:optical-prescription}. The reflected channel is identical, only with surface 13 replaced by a reflection from the first surface of the multi-chroic}\label{fig:layout} 
\end{figure}

The ETSI optical layout is shown in Figure \ref{fig:layout} and prescription in Table \ref{tab:optical-prescription}.

\begin{table}[ht]
\centering
\caption{ETSI Optical Prescription. Radius, Thickness, and Diameter are in millimeters.  The reflected channel is identical, only with surface 13 replaced by a reflection from the first surface of the multi-chroic.}
\label{tab:optical-prescription}
\begin{tabular}{cccccl}
\hline
Surface & Radius   & Thickness & Glass    & Diameter & Comment                  \\
\hline
1       & Infinity & 108.986   &  -       & 72.0     & Telescope Focal Plane    \\
2       & 165.729  & 10.628    &  S-BAL35 & 92.8     & Collimator L1            \\
3       & Infinity & 162.346   &  -       & 92.8     & -                        \\
4       & -54.179  & 7.670     &  N-KZFS2 & 67.5     & Collimator L2            \\
5       & 83.360   & 0.497     &  -       & 67.5     & -                        \\
6       & 79.519   & 28.000    &  S-BAL3  & 73.3     & Collimator L3            \\
7       & -158.561 & 4.130     &  -       & 73.3     & -                        \\
8       & 237.390  & 9.491     &  S-BSM15 & 75.5     & Collimator L4            \\
9       & 78.961   & 1.250     &  -       & 75.5     & -                        \\
10      & 88.455   & 22.002    &  S-FPL55 & 75.5     & Collimator L5            \\
11      & -79.436  & 15.000    &  -       & 75.5     & -                        \\
12      & Infinity & 105.000   &  -       & 30.3     & Shutter                  \\
13      & Infinity & 5.000     &  N-BK7   & 65.0     & Multichroic - 30deg tilt \\
14      & Infinity & 60.000    &  -       & 65.0     & -                        \\
15      & Infinity & 40.000    &  N-SF5   & 75.0     & Prism                    \\
16      & Infinity & 45.000    &  -       & 75.0     & -                        \\
17      & Infinity & 5.000     &  N-BK7   & 88.5     & Clean-up Filter          \\
18      & Infinity & 15.000    &  -       & 88.5     & -                        \\
19      & 87.696   & 21.358    &  S-FPL51 & 81.6     & Camera L1                \\
20      & -112.607 & 3.219     &  -       & 81.6     & -                        \\
21      & -83.695  & 12.324    &  S-LAL13 & 81.6     & Camera L2                \\
22      & 191.915  & 0.530     &  -       & 81.6     & -                        \\
23      & 193.382  & 30.000    &  S-FPL51 & 85.5     & Camera L3                \\
24      & -80.456  & 1.478     &  -       & 85.5     & -                        \\
25      & -78.202  & 7.431     &  S-LAL13 & 76.9     & Camera L4                \\
26      & -119.932 & 5.911     &  -       & 76.9     & -                        \\
27      & 83.321   & 22.000    &  S-FPL55 & 84.1     & Camera L5                \\
28      & -144.369 & 48.949    &  -       & 84.1     & -                        \\
29      & -41.501  & 4.999     &  S-LAH63 & 38.2     & Camera L6                \\
30      & Infinity & 15.000    &  -       & 38.2     & -                        \\
31      & Infinity & 2.500     &  NIFS-V  & 45.6     & Detector Window          \\
32      & Infinity & 4.300     &  -       & 45.6     & -                        \\
33      & Infinity & -         &  -       & 32.0     & Image Plane             
\end{tabular}
\end{table}

\clearpage
\vspace{2ex}\noindent\textbf{Luke Schmidt} is the Yerkes Observatory Project Scientist and was previously an Associate Research Scientist at Texas A\&M University in the Department of Physics \& Astronomy (2015-2024), and Instrument Scientist at the Magdalena Ridge Observatory Interferometer (2013-2015). He received his BS in Physics and Chemistry from Bethel College, KS in 2003, MS in Physics from the New Mexico Institute of Mining and Technology in 2009, and PhD in Physics with Dissertation in Astrophysics from the New Mexico Institute of Mining and Technology in 2012. His research focuses on astronomical instrumentation in the optical and near infrared and includes instruments for imaging, spectroscopy, interferometry, and telescope calibration systems.  He is a member of SPIE.

\vspace{1ex}
\noindent Biographies of the other authors are not available.

\end{document}